\documentclass[twocolumn,prl,superscriptaddress,sort&compress,amsmath,amssymb,floatfix]{revtex4-1}

\DeclareUnicodeCharacter{25CF}{$\bullet$}

\usepackage{graphicx}
\usepackage[dvipsnames]{xcolor}
\usepackage{amssymb}
\usepackage{amsmath}
\usepackage{amsbsy}
\usepackage{color}
\usepackage{ulem}
\usepackage{bm}

\begin{document}
\title{A model for spontaneous SMCs unidirectional motion}
\title{Spontaneous Unidirectional Loop Extrusion Emerges from Structural Symmetry Breaking of SMC-Protein Motion}
\title{Spontaneous Unidirectional Loop Extrusion Emerges from Symmetry Breaking of SMC Extension}

\author{Andrea Bonato}
\thanks{corresponding author, andrea.bonato@strath.ac.uk}
\affiliation{Department of Physics, University of Strathclyde, Glasgow, G4 0NG, UK}
\author{Jae-Won Jang}
\affiliation{Department of Physics and Astronomy, and Institute of Applied Physics, Seoul National University, Seoul, 08826, South Korea}
\author{Kyoung-Wook Moon}
\affiliation{Department of Physics and Astronomy, and Institute of Applied Physics, Seoul National University, Seoul, 08826, South Korea}
\author{Davide Michieletto}
\thanks{co-corresponding author, davide.michieletto@ed.ac.uk}
\affiliation{School of Physics and Astronomy, University of Edinburgh, Peter Guthrie Tait Road, Edinburgh, EH9 3FD, UK}
\affiliation{MRC Human Genetics Unit, Institute of Genetics and Cancer, University of Edinburgh, Edinburgh EH4 2XU, UK}
\author{Je-Kyung Ryu}
\thanks{corresponding author, prof.love@snu.ac.kr}
\affiliation{Department of Physics and Astronomy, and Institute of Applied Physics, Seoul National University, Seoul, 08826, South Korea}

\begin{abstract}
\textbf{DNA loop extrusion is arguably one of the most important players in genome organization. The precise mechanism by which loop extruding factors (LEFs) work is still unresolved and much debated. One of the major open questions in this field is how do LEFs establish and maintain unidirectional motion along DNA. In this paper, we use High-Speed AFM data to show that condensin hinge domain displays a structural, geometric constraint on the angle within which it can extend with respect to the DNA-bound domains. Using computer simulations, we then show that such a geometrical constraint results in a local symmetry breaking and is enough to rectify the extrusion process, yielding unidirectional loop extrusion along DNA. Our work highlights an overlooked geometric aspect of the loop extrusion process that may have a universal impact on SMC function across organisms.
}
 \end{abstract}

\maketitle

DNA loop extrusion by structural maintenance of chromosome (SMC) complexes has emerged as a universal organizing principle for chromosomes~\cite{nasmythCohesinCatenaseSeparate2011,Alipour2012,Sanborn2015a,fudenbergFormationChromosomalDomains2016,Hirano2016,Davidson2021,Xiang2021}. For instance, it is now well established that in eukaryotes, cohesin complexes  are involved in shaping ``topologically associating domains'' (TADs) during interphase~\cite{Vian2018,Rao2017}, while condensin complexes direct the establishment of the cylindrical structure of mitotic chromosomes~\cite{gibcusPathwayMitoticChromosome2018}. 

%DM 
%In spite of this, there is still contrasting evidence regarding the precise topology and mechanics of the loop extrusion process.

\begin{figure}[t!]
	\centering
 \includegraphics[width=0.42\textwidth]{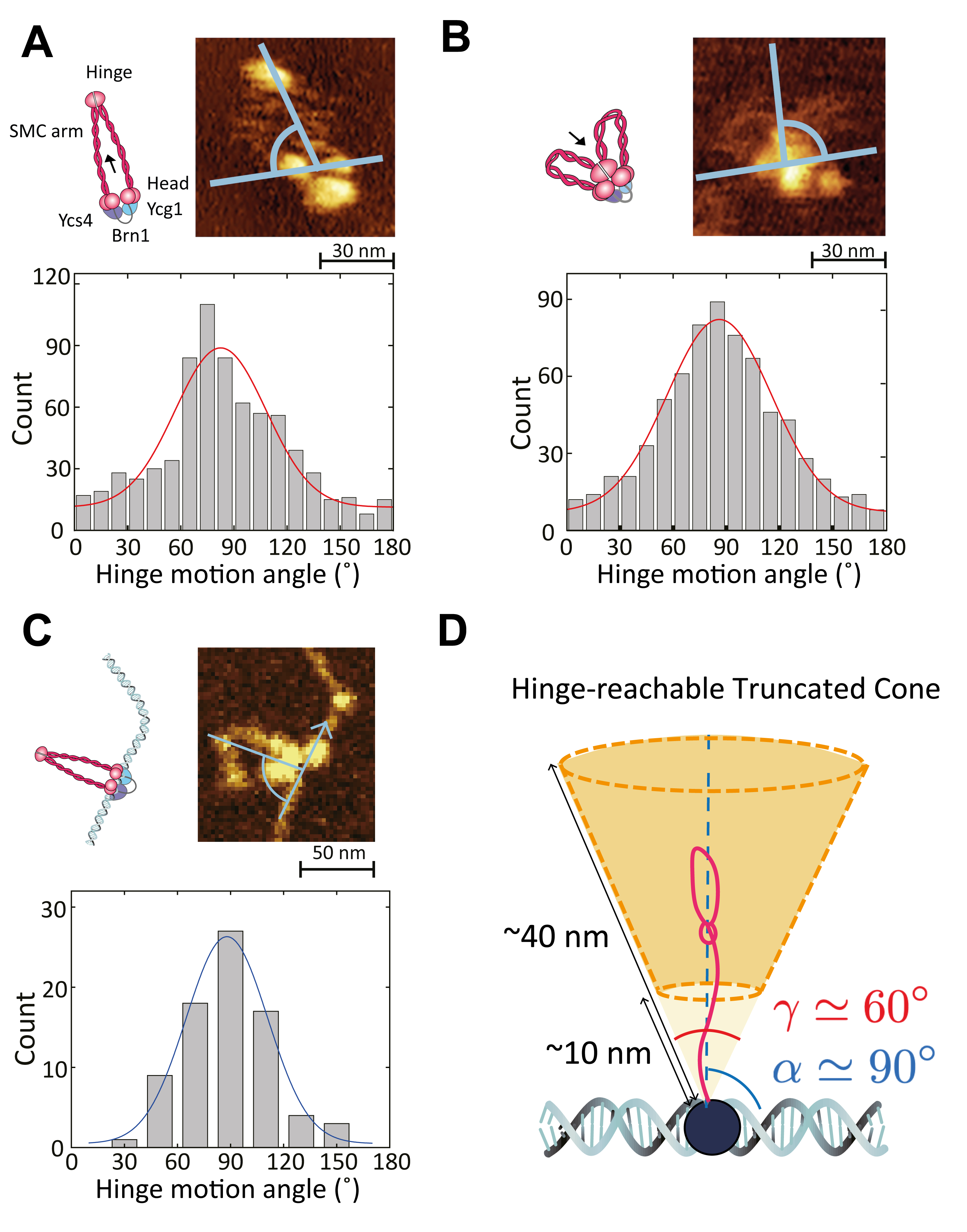}
  \vspace{-0.2 cm}
	\caption{\textbf{Limited solid angle of hinge movement from the position of the SMC heads.} \textbf{a-b.} Angle distributions of hinge-releasing \textbf{a.} and hinge-engaging movements \textbf{b.} respect to the line connecting two heads at the middle of two heads observed by HS AFM (\textit{N} = 727 and 698, respectively). \textbf{c.} Distributions of the angle between hinge and the central position of the globular domain with respect to tangential DNA line at the central positions of the globular domain analysed by dry-AFM images (N = 79). \textbf{d.} Sketch of the truncated cone hinge-reachable region determined by the AFM data.}
  \vspace{-0.5 cm}
	\label{fig:afm}
\end{figure}

%JK suggestion
SMC complexes have a ring-like structure, composed by a SMC dimer and an intrinsically disordered kleisin subunit. The SMC dimer is formed through the so-called ``hinge'' while the kleisin subunit is bound to the ATPase domain of each SMC (Fig.~\ref{fig:afm}a). In the case of yeast condensin, there are putative additional DNA binding sites in the hinge, dimerized heads, Ycg1/Brn1 and Ycs4/Brn1 for DNA anchoring~\cite{Shaltiel2022}. Despite the wealth of structural data, there is still contrasting evidence regarding the precise topology and mechanics of loop extrusion process.

Among the most debated features of loop extrusion are the motoring action of the hinge and the origin of the unidirectional motion~\cite{Higashi2021,Nomidis2021,Takaki2020,Barth2023}.
Recent structural studies have suggested that SMC uses conformational changes between a hinge-released state -- where the hinge is extended away from the heads -- and a hinge-engaged state -- where the hinge is in proximity of the heads, in order to drive the motion in a ``scrunching'', and ATP-dependent, fashion~\cite{Ryu2020,Bauer2021,Xiang2021} (Fig.~\ref{fig:afm}a, b). The scrunching model predicts that following dimerisation of the SMC heads (due to ATP-binding), the coiled-coil arms fold to bring the hinge closer to the heads. Following ATP-hydrolysis, the heads are released and the hinge extends again~\cite{Ryu2020,Eeftens2016,Takaki2020}. During this step, the positively charged extended hinge may search for a 3D proximal (but not necessarily 1D contiguous) DNA segment to grab and to subsequently bring close to the heads in the following ATP-binding step~\cite{bonatoThreedimensionalLoopExtrusion2021,Takaki2020}. 
Whilst this model can elegantly explain the bypassing of other SMCs~\cite{Kim2020,Brandao2021} and large roadblocks~\cite{Brandao2019,Pradhan2021}, it cannot explain unidirectionality. Indeed, during the search step, there is no guarantee that the hinge will grab onto a DNA segment ahead of the one that was reeled in the extruded loop in the previous ATP cycle. Thus, even within the scrunching model explaining the unidirectional motion of LEFs remains an outstanding problem in the field of SMC-driven DNA organisation.

To gain a better understanding of the origin of the unidirectional motion, we use High-Speed AFM (HS AFM) on yeast condensin~\cite{Ryu2020}. We observed that the hinge domain typically extends in an orthogonal direction from the DNA-bound globular head domains, and we then hypothesised that this local (angular) symmetry breaking due to the structure of condensin could generate a geometric bias in the grabbing of new DNA segments. Thus, we computationally implemented a loop extrusion process with such geometric constrain, by modelling a 3D search constrained to lie within a solid angle, and we discovered that it spontaneously displayed rectified, unidirectional extrusion. Our findings suggest that a geometric symmetry-breaking mechanism, without the need to impose explicit topological constraints between the SMC complex and extruded DNA segment, could underlie the emergence of rectified, unidirectional loop extrusion by SMC proteins. 

\paragraph{AFM reveals a preferred angle for hinge extension. -- } We started from the hypothesis that the structure of SMC itself may present some geometric restrictions on its allowed conformations. Specifically, we argued that the hinge, connected to dynamic coiled-coil arms~\cite{Eeftens2016}, largely determines the search of the DNA segment to be captured and reeled in the extruded loop. For this reason, we wanted to measure the possible geometric conformations assumed by a LEF, in order to quantify any restriction on the motion of the hinge. To do this, we performed and analysed High-Speed AFM (HS AFM) images of yeast condensin complex to obtain the typical position of the hinge with respect to the globular head domains~\cite{Ryu2020} (Fig.~\ref{fig:afm}). 
From the HS AFM movies we could identify two distinct globular domains: the hinge and the heads, linked together by semi-flexible arms (Fig.~\ref{fig:afm}a). Furthermore, we were able to distinguish hinge-released and hinge-engaged states by measuring the distance of the hinge from the heads (see Fig.~\ref{fig:afm}a-b and SI Fig.~S4). Based on these, we measured the angles of hinge-releasing and hinge-engaging states, with respect to the line connecting two heads and from the middle point of two heads (Fig.~\ref{fig:afm}a-b). In addition, to compare these angles with the anchoring angle of condensin with respect to the DNA, we analyzed dry-AFM images and measured the angle of hinge extension with respect to the DNA-bound head domains (Fig.~\ref{fig:afm}c).  

We discovered that the hinge is most often extended orthogonally to the line joining the SMC heads (Fig.~\ref{fig:afm}a-b). Even in absence of DNA, we measured that the angle distribution at which the hinge is extended and retracted is normally distributed for both hinge releasing ($83 \pm 26^\circ$) and engaging ($86 \pm 30^\circ$) steps.
In addition, dry-AFM images of condensins that bound to DNA through head domains also showed vertical angle of the hinge-head line respect to DNA tangential direction ($88 \pm 23^\circ$) (Fig.~\ref{fig:afm}c). 

It is important to notice that our measurements are done on condensin complex absorbed on mica, i.e., in 2D. Thus, we argue that the angle distribution of the hinge extension would define a solid angle when the complex is allowed to move in 3D. Indeed, our results suggest that the hinge releases and engages orthogonally to both heads and DNA, forming a solid angle $\Omega$ defined by the width of the Gaussian distribution, $\gamma \simeq 60^\circ$ (see Fig.~\ref{fig:afm}d).

Based on the AFM results, we defined a hinge-reachable region for the scrunching model as a truncated cone with the estimated solid angle $\Omega$ (Fig.~\ref{fig:afm}d). The maximum hinge extension is determined from the distribution of hinge-head distances in the hinge-released state ($\simeq 40$ nm) while the minimum hinge extension is obtained as the hinge-head distance in the hinge-engaged state~\cite{Ryu2020} ($\simeq 10$ nm, see SI Fig.~S4).
 
\begin{figure*}[t!]
	\centering
	\includegraphics[width=0.9\textwidth]{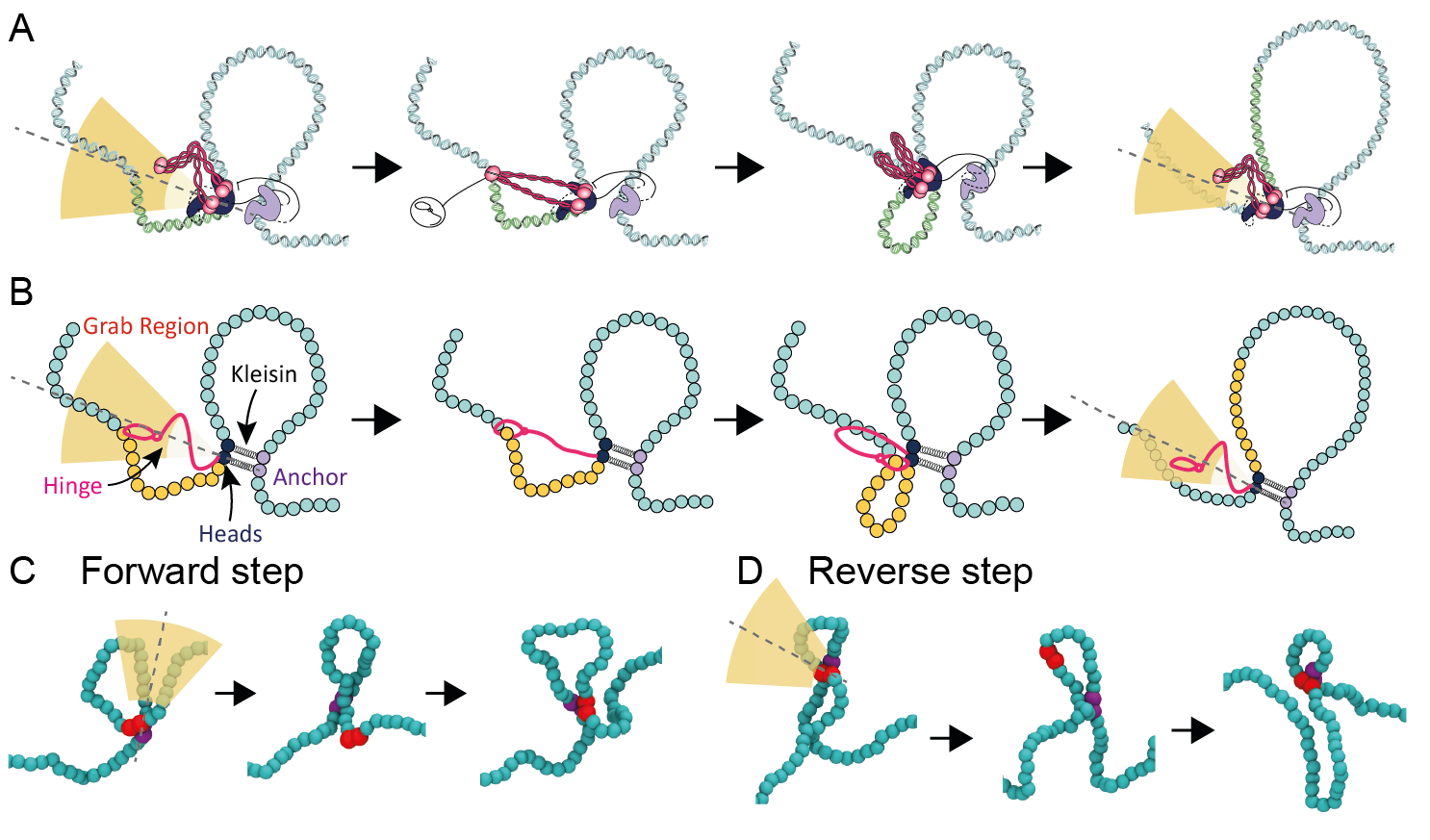}
  \vspace{-0.6 cm}
	\caption{ \textbf{A model for loop extrusion with geometric constraints on hinge extension.} \textbf{a.} Schematics of a loop extrusion model where the hinge (red) is restricted to search for a DNA segment within a truncated cone. The segment is then bound to the SMC heads and reeled in the extruded loop, while the hinge returns to the search position. Throughout this process, one DNA segment is trapped at the anchor bound to the kleisin (purple). \textbf{b} Implementation of the model on a coarse-grained bead-spring polymer, where the heads and anchor are denoted with red and purple beads, respectively. The hinge is not explicitly modelled with a bead, but is accounted for by the geometrically restricted search region (yellow shaded area). The simulated loop extrusion displays both (\textbf{c}) forward and (\textbf{d}) backward steps. 
    }
     \vspace{-0.4 cm}
	\label{fig:model}
\end{figure*}

\paragraph{A LEF model with geometrically biased 3D search. -- }

In light of our HS AFM measurements, we propose a new geometrically-constrained scrunching model as follows: first, Ycg1/Brn1 anchors DNA with a ``safety belt'' mechanism~\cite{Kschonsak2017} then the heads/Ycs4 connect the anchors to other part of DNA to extrude a loop (Fig.~\ref{fig:model}a). The motor action of the hinge is limited by the angle distribution we found in Fig.~\ref{fig:afm}, and can grab DNA segments through the hinge, by extending the SMC arms at a fixed angle $\alpha \simeq 90^\circ$ and with a certain width $\gamma \simeq 60^\circ$ from the orientation of the bound DNA. After that the hinge grabs onto a new DNA segment, the ATP-binding-induced conformational change brings the grabbed DNA close to the heads/Ycs4 subunits. Finally, after ATP-hydrolysis, the heads/Ycs4 bind to the new DNA segment -- thereby extending the extruded loop -- and the hinge is then free to target a new DNA segment for the next round of DNA-loop extrusion (Fig.~\ref{fig:model}a). 

To simulate this model, we implemented a coarse-grained loop extrusion process with a geometric constraint on the region that can be reached by the hinge. Specifically, we account for the connectivity of the anchor (YcgI) to the heads (Smc2 and Smc4) via the kleisin subunit as beads connected by a harmonic bond; additionally, we impose that the search of the segment to reel in the extruded loop is to be performed within a solid angle around an axis orthogonal to the tangent of the SMC-bound DNA (Fig.~\ref{fig:model}b). When a segment of the coarse-grained polymer falls within a truncated cone formed by the solid angle $\Omega$ and within a certain Euclidean distance (10 and 40 nm) the position of the harmonic bond between anchor and heads is updated to grab onto the anchor and such new segment (see Fig.~\ref{fig:model}b). In turn, the harmonic spring connecting anchor and heads is temporarily extended and brings the new segment close to the anchor. Finally, the new segment is identified with the new position of the heads, the anchor remains at its original position, and the hinge is then returned free to search for a new segment to grab onto (again within the solid angle $\Omega$). This process defines a full ATP-cycle and involves a 3D search of proximal DNA segments with a geometric constraint but no strict topological requirements.

\begin{figure*}[t!]
	\centering
	\includegraphics[width=0.95\textwidth]{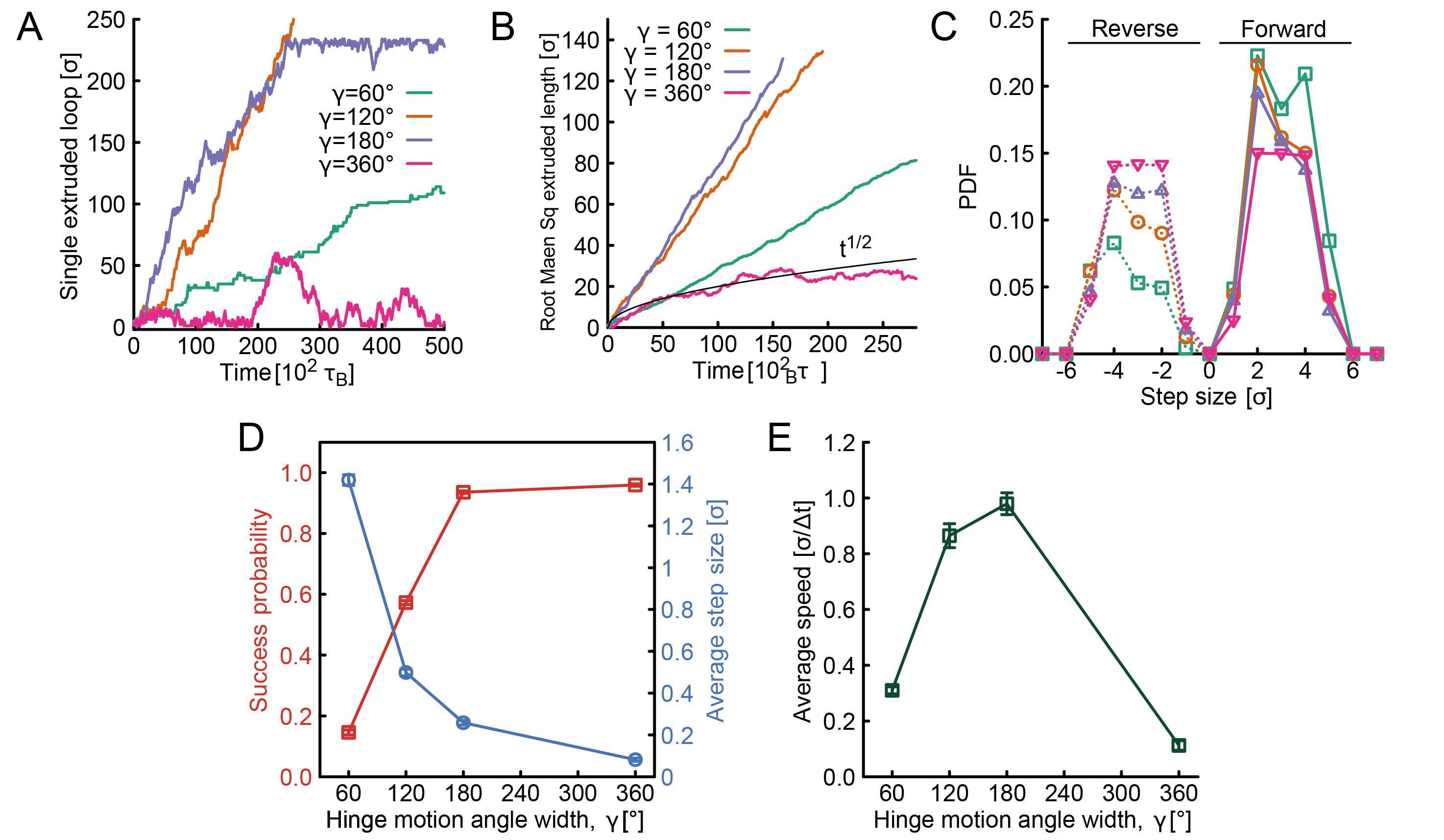}
 \vspace{-0.2 cm}
	\caption{\textbf{Unidirectional motion emerges from symmetry breaking.} \textbf{a.} Individual traces of simulated extruded loops as a function of time and with different hinge-search angles. \textbf{b.} Root mean squared extruded length as a function of time and for different hinge-search angles. The case where the search is allowed to occur on the full sphere $\gamma=360^\circ$ yields a random walk scaling as $t^{1/2}$. \textbf{c.} Probability of step size as a function of the search angle. The narrower the angle, the more rectified the extrusion, i.e. the larger the forward/reverse ratio. \textbf{d.} Plot of the success probability and average step sizes as a function of the hinge search angle. These two quantities display a trade-off which yield an optimum of the extrusion velocity at $\gamma = 180^\circ$ search angles, as shown in \textbf{e}.  
     }
     \vspace{-0.4 cm}
	\label{fig:unidirectional}
\end{figure*}

The important difference of our work from previous models of loop extrusion~\cite{Sanborn2015a,fudenbergFormationChromosomalDomains2016,Alipour2012,baniganLimitsChromosomeCompaction2019,Orlandini2019,bonatoThreedimensionalLoopExtrusion2021,Takaki2020} is that we do not impose the directionality of the extrusion \textit{a priori}. The hinge can grab any segment ahead, or behind, the current 1D position of the heads. In fact, in our simulations we can observe backward extrusion steps, where the newly grabbed segment is \textit{inside} the extruded loop, as shown in Fig.~\ref{fig:model}c, d. This move yields a reduction in the total length of the extruded loop and is also observed in experiments~\cite{Ryu2022}. In addition, our model accounts for an additional spring to mimic the presence of the disordered kleisins attaching the anchor YcgI to the SMC heads (SI). This additional spring constrains the relative rotation of the DNA segments bound to the heads and the anchor. This rotational constraints cannot otherwise be imposed if two beads are connected by a single spring~\footnote{An alternative way to implement this constraints could be to impose a dihedral or angular potential between the DNA segments bound to heads and anchor. We expect to obtain similar results with this alternative implementation.}. The rotational constraint within the DNA-bound SMC complex is evident from structural cryo-EM data~\cite{Shaltiel2022}, where the bound DNA segments sit tightly within DNA-binding pockets near the heads, and that their juxtaposition within the SMC structure assume a well-defined angle~\cite{Shaltiel2022}; this implies that SMC-bound DNA segments are likely to be restricted in their relative rotation. 

Using this model, we performed a simulated loop extrusion on a bead-spring polymer with $N=400$ beads of size $\sigma = 10$ nm. We then tracked the position of the anchor and heads, and defined an oriented extruded loop length as $l = n_a - n_h$, where $n_a$ and $n_h$ are the positions of the anchor and the heads; we discovered that, strikingly, the LEFs display growing loops with a clear sign of unidirectional motion. Since we do not hard-code directionality within the model, the LEFs in different replicas will start extruding in different directions. The interesting finding is that they display a tendency to maintain a rectified, unidirectional motion, once that the ``left-right'' symmetry has been broken (Fig.~\ref{fig:unidirectional}a). 

To understand how this spontaneous rectification is due to the broken symmetry in the search step, we performed simulations with wider search angles up to restoring the full spherically symmetric search $\gamma = 360^\circ$. In the symmetric case, we do not observe unidirectional motion, instead we find that extruded loops shrink back, with a behaviour similar to a random walk (see Fig.~\ref{fig:unidirectional}a). To further characterise this process we took the root mean squared extruded length $\langle l \rangle = \langle \left[ n_a - n_h \right]^2 \rangle^{1/2}$ and indeed found that the spherically symmetric case displayed a scaling $\langle l \rangle \sim t^{1/2}$, in line with a simple random walk (see Fig.~\ref{fig:unidirectional}b). Interestingly, we also noted that the cases with $60^\circ < \gamma < 360^\circ$ displayed faster linear growth than the case with $\gamma = 60^\circ$. Despite this, the distribution of step sizes clearly indicate that the case $\gamma = 60^\circ$ is the one that benefits from the greater rectification, i.e. the ratio forward/reverse steps is the largest. In other words, wider angles increase the probability of shorter and backward steps (Fig.~\ref{fig:unidirectional}c). In turn, this implies that the average step size -- defined as $s = \Sigma_i [sign(i) S_i]/N$, where $S_i$ is the $i$-th step size -- is typically smaller for wider angles. The largest probability of large steps $\sim 50$ nm, which is in line with experiments~\cite{Ryu2022}, was seen for $\gamma = 60^\circ$. Thus, to understand why wider search angles yield faster extrusion in our simulations, we compute the success probability of making a step. Indeed, in our algorithm we impose that the LEF does not make a step if, in a given simulation time, there are no DNA beads that satisfy the search criterion. This readily implies that narrower search angles yield lower success rates (Fig.~\ref{fig:unidirectional}d). The opposite trends of success rate (increasing with $\gamma$) and average step size (decreasing with $\gamma$) yield a trade-off (Fig.~\ref{fig:unidirectional}d) that naturally leads to an optimum in velocity around $\gamma \simeq 180^\circ$. However, we argue that condensin may employ a narrow angle to optimize the search process within a crowded DNA environment, within which the success probability would be generally larger.  

\paragraph{Conclusions. --} 
Motivated by AFM data, in this Letter we have provided experimental evidence that the structure of yeast condensin favours certain geometric conformations where the hinge is extended perpendicularly to the local direction of the heads-bound DNA segment. We have also quantified the width of the search angle, $\gamma = 60^\circ$ and computationally demonstrated that by imposing this geometric constraint, loop extrusion can be spontaneously rectified. Interestingly, we find that the narrower the search angle, the larger the typical the step size and the more unidirectional the extrusion, but also the more likely to fail to find a DNA segment to grab in a given time. This trade-off yields an optimum extrusion speed at a predicted angle of $\gamma = 180^\circ$. 

We argue that the emergence of spontaneously rectified, unidirectional extrusion is due to a combination of local deformations of the underlying DNA and the geometric constraint on the hinge-mediated 3D search. Our findings ought to be relevant to other SMC protein complexes, as long as their structure impose a geometric constraint on the conformational space of the DNA:protein complex. Additionally, they can reconcile a range of recent findings, e.g. bypassing of roadblocks, Z-loops, and also the pinching of a negatively supercoiled loop during the hinge-grabbing step~\cite{rocapinches2023}, whilst also explaining the unidirectional extrusion. 
Our hypothesis could be tested by mutating the SMC coiled-coil arms to be more flexible or rigid, thereby directly increasing or reducing the search angle $\gamma$. Overall, we argue that our findings contribute to highlight a largely overlooked aspect of loop extrusion that ought to be relevant for the function of generic SMC complexes.

\paragraph{Acknowledgements. --}
DM acknowledges the Royal Society and the European Research Council (grant agreement No 947918, TAP) for funding. The authors also acknowledge the contribution of the COST Action Eutopia, CA17139. J.-K.R. acknowledges the Institute of Applied Physics of Seoul National University, Creative-Pioneering Researchers Program through Seoul National University (Project Number 3348-20230013), the Brain Korea 21 Four Project grant funded by the Korean Ministry of Education, Samsung Electronics Co., Ltd. (Project Number A0426-20220109), and the National Research Foundation of Korea (Project Number 0409-20230237, 0409-20230171, 0409-20230219).

\vspace{-0.2 cm}

\bibliographystyle{apsrev4-1}
\bibliography{biblio}

\vspace{-0.2 cm}

%\section{Data availability}
%\dm{Datasets used to generate the figures in this article have been deposited at xx}

%\section{Code availability}
%\dm{Our codes are deposted at xx}.

\end{document}

\title{Spontaneous Unidirectional Loop Extrusion Emerges from Symmetry Breaking of SMC Extension}

\author{Andrea Bonato}
\thanks{corresponding author, andrea.bonato@strath.ac.uk}
\affiliation{Department of Physics, University of Strathclyde, Glasgow, G4 0NG, UK}
\author{Jae-Won Jang}
\affiliation{Department of Physics and Astronomy, and Institute of Applied Physics, Seoul National University, Seoul, 08826, South Korea}
\author{Kyoung-Wook Moon}
\affiliation{Department of Physics and Astronomy, and Institute of Applied Physics, Seoul National University, Seoul, 08826, South Korea}
\author{Davide Michieletto}
\thanks{co-corresponding author, davide.michieletto@ed.ac.uk}
\affiliation{School of Physics and Astronomy, University of Edinburgh, Peter Guthrie Tait Road, Edinburgh, EH9 3FD, UK}
\affiliation{MRC Human Genetics Unit, Institute of Genetics and Cancer, University of Edinburgh, Edinburgh EH4 2XU, UK}
\author{Je-Kyung Ryu}
\thanks{corresponding author, prof.love@snu.ac.kr}
\affiliation{Department of Physics and Astronomy, and Institute of Applied Physics, Seoul National University, Seoul, 08826, South Korea}

\setcounter{figure}{0}
\renewcommand{\figurename}{Fig.}
\renewcommand{\thefigure}{S\arabic{figure}}

\maketitle

\section{Model}

We perform Molecular Dynamics (MD) simulations of a loop extruding factor (LEF) operating on linear DNA in the NVT ensamble (constant temperature). DNA is modelled as a bead-spring polymer of 400 beads of size $\sigma = 10$ nm, for a total contour length of $4\ \mu$m. To account for excluded volume interactions, non-bonded beads repel each other according to the  
the Weeks-Chandler-Andersen (WCA) potential:
\begin{equation}\label{UWCA}
U_{WCA}(r) = \left\{\
\begin{array}{lcl}
4\epsilon \left [ \left ( \frac{\sigma}{r}\right )^{12} - \left (\frac{\sigma}{r} \right )^6 \right ] + \epsilon & r \leq r_c\\
0 & r > r_c
\end{array} \right. \, ,
\end{equation}
where $r$ is the separation between the beads and $r_c=2^{1/6}\sigma$. $\epsilon = k_B T$ is the energy unit.
Neighbouring beads are held together by finite-extension-nonlinear-elastic (FENE) bonds. Their interaction is governed by the potential:
\begin{equation}\label{UFENE}
\begin{split}
U_{\rm FENE}(r) = & U_{WCA}(r)+ \\
&+\left\{
\begin{array}{lcl}
-0.5kR_0^2 \ln\left(1-(r / R_0)^2\right) & \ r\le R_0 \\ \infty & \
r> R_0 &
\end{array} \right. \, ,
\end{split}
\end{equation}
where $r$ is the distance between the the beads, $k = 30\epsilon/\sigma^2$ is the spring constant and $R_{0}=1.5\sigma$ is the maximum extension of the bond.
DNA is a semi-flexible polymer with persistence length $l_p \sim 50$ nm ($5\sigma$); we include stiffness in the model by adding a Kratky-Porod energy penalty, acting on triplets of consecutive beads:
\begin{equation}\label{UKP}
U_{KP}(\theta) = \dfrac{k_BT l_p}{\sigma}[1-\cos(\theta)],
\end{equation}
where $\theta$ is the angle formed by the three beads of a given triplet.

\begin{figure}[t!]
	\centering
	\includegraphics[width=0.35\textwidth]{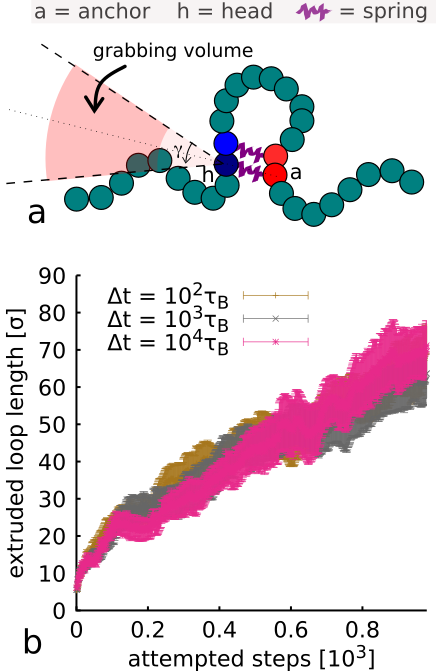}
    %\vspace{-1cm}
	\caption{\textbf{a} Sketch of the loop extrusion model. The LEF is modelled as two harmonic springs holding together a DNA loop and the position of the head, at the base of the loop, is updated (i.e. a new head is selected) every $\Delta t$ timesteps. Candidate head beads are searched within the grabbing volume at every update.  \textbf{b} Average extruded loop length as a function of the number of attempted extrusion steps for $\Delta t = 10^2, 10^3$ and $10^4$ $\tau_B$. }
 %Results from simulations of the one spring model (see section One spring vs two springs).}
	\label{fig:model}
\end{figure}

Each LEF is modelled as a pair of springs bringing together two short DNA segments made of pairs of neighbouring beads (see Fig.~\ref{fig:model}a). This is practically achieved by using two harmonic bonds, described by the potential:
\begin{equation}\label{harmonic}
U_{H}(r) = k\left(r-r_0\right)^2,
\end{equation}
where $r$ is the distance between the bonded beads, $r_0=1.6\sigma$ is the resting distance and $k=5\epsilon$ is the elastic constant. To integrate the equations of motion we use the LAMMPS package~\cite{LAMMPS}. The integration timestep is set to $dt = 0.01\tau_B$, where $\tau_B$ is the Brownian time of a DNA bead.

To load a LEF on DNA, first we randomly select two DNA beads, which must be proximate in 3d (distance $<4\sigma$), and representing the position along DNA of the anchor and the head of the LEF. Then, the two harmonic bonds modelling the LEF are added: one connects the the head to the anchor, the other spring connects the two beads upstream (i.e. inside the extruded loop, see Fig.~\ref{fig:model}a). Note that to do this, the two randomly chosen beads must be at least 2 beads apart. To practically simplify the addition and removal of the harmonic bonds, we require them to be at least 3 beads apart. 

Once loaded, we attempt to update the position of the springs at regular intervals $\Delta t$. The update rule for the extrusion follows these rules: let us suppose that a LEF loaded at time $t_0$ attempts to take a step at time $t=t_0+n\Delta t$, $n \in \mathbb{N}$, and let us call $\textbf{r}_h(t)$ and $\textbf{r}_a(t)$ the position of the head and anchor at time $t$, and $\textbf{dir}(t) = \textbf{r}_h(t) -\textbf{r}_a(t)$ the vector going from the anchor to the head. First, all DNA beads for which the position at time $t$, $\textbf{r}(t)$, satisfies 
$$\frac{\textbf{dir}(t) \cdot \textbf{pos}(t)}{|\textbf{dir}(t)||\textbf{pos}(t)|} \leq \cos(\gamma/2), $$ 
where $\textbf{pos}(t) = \textbf{r}(t) - \textbf{r}_h(t)$ and $\gamma$ is the grabbing angle, and $\sigma \leq |\textbf{pos}(t)| \leq 4\sigma$, are identified. Then one of these beads (if any) is randomly selected (see Fig.~\ref{fig:model}a) and the current bead corresponding to the heads position is deleted and two new bonds, connecting the anchor to the new head, and their upstream neighbours are added to the DNA. If none of the DNA beads at a given timestep satisfy the conditions above, the bonds are left unchanged. Note that, whereas the head can be updated at every step, the anchor never changes. Additionally, we do not impose a direction a priori, these selection rules can chose any bead up- or downstream of the current position of the head. For simplicity, to follow the growth of the extruded loops without disruptions, we also disallow big jumps in extruded length by excluding from the selection rule beads which are farther than 5 beads (in 1D) from the current head position. In other words, we set what we previously defined inter-strand grabbing probability used in Ref.~\cite{bonatoThreedimensionalLoopExtrusion2021} to 0. 

We fix $\Delta t = 10^2 \tau_B$ and run each simulation for $10^5 \tau_B$. Fig.~\ref{fig:model}b verifies that the value of $\Delta t$ we chose is large enough to allow the local equilibration of DNA, and that, for large enough $\Delta t$, regardless of its value, the extrusion speed is a function of the number of attempted steps.

\begin{figure}[t!]
	\centering
	\includegraphics[width=0.5\textwidth]{figures/Suppl_1s_vs_2s.png}
    %\vspace{-1cm}
	\caption{\textbf{One spring vs two springs.} \textbf{a} Sketch of the extrusion model with one harmonic spring. \textbf{bc} Comparison of individual (\textbf{b}) and average (\textbf{c}) extruded length as a function of time as predicted by the two springs model and the one spring model. \textbf{d} Step size distributions for the one spring and two springs models.}
	\label{fig:1svs2s}
\end{figure}

\section{One spring vs two springs}

To emphasise how the extrusion unidirectionality emerges spontaneously from breaking the local symmetry in our simulated system, here we investigate what happens if we remove one of the two harmonic bonds stabilising the extruded loop, precisely the spring connecting the upstream neighbours of the head and the anchor (see Fig.~\ref{fig:1svs2s}a). Note that, to compensate for the removal of one spring, for this set of simulations we double the spring constant ($k=10\epsilon$) of the remaining harmonic bond. Fig.~\ref{fig:1svs2s} shows that the extrusion process is much more efficient with two springs, as quantified both by the extruded length as a function of time (Fig.~\ref{fig:1svs2s}bc) and the distributions of forward and reverse steps (Fig.~\ref{fig:1svs2s}d). We argue that this difference is due to the fact that adding a second spring inside the loop sharpens the local conformational asymmetry of the DNA inside with respect to outside the extruded loop.

\section{Rectification is enhanced on stretched DNA}

\begin{figure}[h!]
	\centering
	\includegraphics[width=0.4\textwidth]{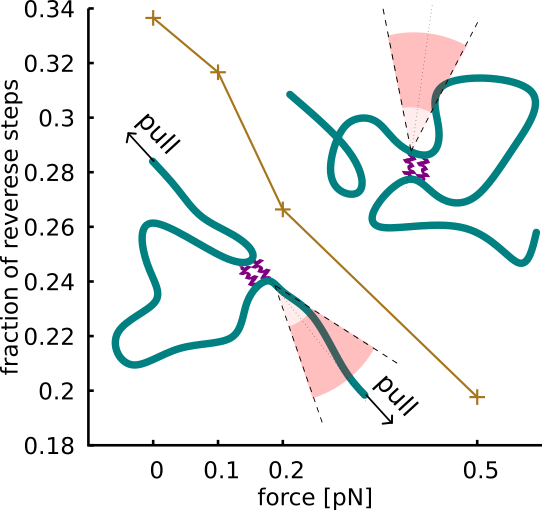}
    %\vspace{-1cm}
	\caption{\textbf{Directionality for stretched DNA.} Fraction of reverse steps for different pulling forces. The inset sketches illustrate a LEF operating on a substrate with free ends (top right) and one under tension (bottom left).}
	\label{fig:pulling}
\end{figure}

In this section we explore how pulling the DNA ends with a constant force, relevant, for instance, in the case of tethered DNA, affects the extrusion directionality.
Fig.~\ref{fig:pulling} shows that, as the stretching force increases, the loop extrusion process becomes more rectified, with the LEF taking fewer reverse steps. Due to the difference in the tension experienced by the DNA portions outside and inside the extruded loop, pulling the ends is an effective way of strengthening the local conformational asymmetry of the DNA close to the extruded loop boundaries, thus helping the LEF keeping its direction.

\section{Condensin purification}
We used the same protocol as previously reported~\cite{Ryu2020} for the purification of \textit{S. cerevisiae} condensin holocomplexes with all the subunits. 

\section{Dry-AFM sample preparation and imaging}
In order to visualize DNA structures, bound by condensin using AFM, we incubated $\lambda$-DNA (D1501, Promega) at a concentration of 3 ng/$\mu$l with condensin at concentration of 5 nM in an Eppendorf tube. This mixture was incubated for 10 minutes to facilitate the interaction between condensin and DNA. Afterward, 1 mM ATP was added, and the sample was further incubated for an additional minute. The resulting solution was then placed on polylysine-treated mica, with a concentration of 0.00001\% (wt/vol), for 20 seconds. The sample-coated mica was rinsed using 3 ml of MilliQ water, followed by drying using N2 gas.
The AFM measurements were conducted in ambient air utilizing a Bruker Multimode AFM equipped with a Nanoscope V controller and Nanoscope version 9.2 software. Bruker ScanAsyst-Air-HR cantilevers with a nominal stiffness and tip radius of 0.4 N/m and 2 nm was used. The imaging technique employed was PeakForce Tapping, characterized by an 8-kHz oscillation frequency and a peak force setpoint below 100 pN. 
The angle of condensin anchoring to DNA between the line that connecting between hinge and the center of globular domain of condensin bound to DNA and the DNA tangential direction at the center of the globular domain bound to DNA. 

\section{Liquid-phase HS AFM imaging}
Wild-type condensin holocomplex (at      concentration of 2 nM) was deposited onto a freshly cleaved mica surface using a buffer composed of 20 mM TRIS-HCl (pH 7.5), 50 mM NaCl, 2.5 mM MgCl2, 2.5 mM DTT, along with 1 mM ATP. After 10 seconds deposition, the sample surface was rinsed with the same buffer. The condensin sample was imaged using the HS AFM developed by RIBM. This was achieved by using either Nanoworld SD-S-USC-f1.2-k0.15 cantilevers (with a tip radius of 2 nm, spring constant k = 0.15 N/m, and frequency f = 1.2 MHz). During the imaging process, a scan size of 100 nm $\times$ 100 nm was typically used, along with 100–150 scan lines and frame rates ranging from 1 to 10 Hz. The obtained data were processed using a Python script to reconstruct both movie files and images. Within this framework, a Python script was used to measure the angle of hinge-releasing and hinge-engaging movements. To define the transition state between hinge-releasing and hinge-engaging, we measured hinge-head distance and found the time points that showed stepwise increase or decrease of hinge-head distance by applying step-finding algorithm (Supplementary Fig.~S4). At these time points, we measured the hinge-movement angles between the line of two consecutive hinge positions and the line that connecting two heads. 

\begin{figure*}[b!]
	\centering
	\includegraphics[width=0.95\textwidth]{figures/Suppl_Figure1_Directionality_Sketch_v2-01.png}
	\caption{\textbf{Real-time traces of hinge-motion of yeast condensin holocomplex.}  (A) Cartoon of condensin holocomplex. Hinge-head distance was measured for every HS AFM frame. (B) Real-time traces of hinge-head distance changes. To measure the angles of hinge movement respect to the head-head line, the vector constructed by the hinge movement at two consecutive frames where stepwise hinge motion is observed. To find the timepoint when stepwise hinge motion is observed, we applied an automatic step-finding algorithm previously developed in Ref.~\cite{Loeff2021}. (C) Two gaussian distributions of hinge-head distances (N = 3,484 frames, mean $\pm$ SD = 10.8 $\pm$ 3.4 and 32.1 $\pm$ 11 nm). These distributions define the hinge-reachable truncated-cone shaped region. }
	\label{fig:afm}
\end{figure*}

\bibliographystyle{apsrev4-1}
\bibliography{biblio}